\documentclass[prl,twocolumn,showpacs, tightenlines]{revtex4}
\usepackage{graphicx}
\usepackage{amssymb}

\newcommand{\nn}{\nonumber}

\newcommand{\be}{\begin{equation}}
\newcommand{\ee}{\end{equation}}
\newcommand{\bea}{\begin{eqnarray}}
\newcommand{\eea}{\end{eqnarray}}

\begin{document}

\title{Geometrical spin dephasing in quantum dots}

\author{Pablo San-Jose$^1$, Gergely Zarand$^{1,2}$, Alexander Shnirman$^1$,
and Gerd Sch\"on$^1$} \affiliation{ $^1$ Institut f\"{u}r
Theoretische Festk\"{o}rperphysik and DFG-Center
    for Functional Nanostructures (CFN), Universit\"{a}t Karlsruhe,
    D-76128 Karlsruhe, Germany.\\
$^2$ Institute of Physics, Technical University Budapest, Budapest,
H-1521, Hungary. }
\date{\today}

\begin{abstract}
We study spin-orbit mediated relaxation and dephasing of electron
spins in quantum dots. We show that higher order contributions
provide a relaxation mechanism that dominates for low magnetic
fields and is of geometrical origin. In the low-field limit
relaxation is dominated by coupling to electron-hole excitations and
possibly $1/f$ noise rather than phonons.
\end{abstract}
 \maketitle

Recent experiments \cite{Elzerman04,Petta05} demonstrate that spins
of single electrons confined in quantum dot structures can be
manipulated in a quantum coherent way, thus opening exciting
perspectives for  quantum information processing \cite{Loss98}. Thus
a thorough understanding of spin relaxation and decoherence
processes is crucial. This requires identifying the sources of
fluctuations and the mechanisms how they couple to the spins, as
well as analyzing the non-equilibrium decay laws. So far, two major
mechanisms of dephasing have been identified: One is the hyperfine
coupling to randomly oriented {\em nuclear spins}
\cite{Khaetskii02}. While in a free induction decay this leads to
fast dephasing, on a time scale of order $\sim 10$ ns, the effect of
the quasi-static nuclear field can be eliminated largely by
spin-echo techniques, bringing the dephasing time into the range of
$\sim 1 \mu$s \cite{Petta05}. The second mechanism is the coupling
to (piezoelectric) phonons in the presence of {\em spin-orbit}
interaction~\cite{Khaetskii01,Woods02,Golovach04}. Phonons create a
fluctuating electric field acting on the electrons' orbital degrees
of freedom, which couples via the spin-orbit interaction to the
spin. If time reversal symmetry is broken by a magnetic field $B$
the process leads to spin relaxation. This mechanism is important in
sufficiently strong fields~\cite{Khaetskii01}, however, as usually
described in the literature, it is ineffective in vanishing fields
for electrons in the ground state doublet.

In this paper we study, what processes destroy the spin
coherence in vanishing or low magnetic fields.  We show that higher
order (e.g., two-phonon) virtual processes, usually neglected in the
literature, provide a relaxation mechanism that {\em persists} as
$B\to 0$. These relaxation processes are of {\em geometrical} origin
and related to the diffusion of the Berry phase.
Berry phase emerging from the spin-orbit interaction
was introduced in Ref.~\cite{Aronov_Lyanda}.
A different
phenomenon also called {\it geometric dephasing} was discussed in
Ref.~\cite{Whitney05}. The processes we consider are the analogues
of Elliott's spin relaxation in bulk semiconductors and
metals~\cite{Elliott54}, for which a geometrical interpretation has recently
been given in Ref.~\cite{Serebrennikov04}. Processes of this type have
also been studied in the context of phonon-induced relaxation in
electron spin resonance experiments~\cite{Abrahams57}, and were
found to lead to a relaxation even as  $B\to 0$. However, the
geometric origin of the mechanism has not been revealed. Moreover,
the truncation of the Hilbert space to two orbital levels, used
there, is insufficient, since amplitude cancelations from higher
orbitals prove to be crucial.

Moreover, we observe that spin relaxation is induced by
{\em any} kind of electric field fluctuations, not merely by
phonons. In fact, in order to confine, control, and measure the
electron one attaches electrodes and quantum point contacts to the
qubit~\cite{Petta05}. They produce {\em Ohmic} fluctuations with
dominant spectral weight at low frequencies. For typical quantum
dots we find that such Ohmic fluctuations provide the dominant mechanism for
Berry-phase dephasing at $B=0$, and they also
provide  the leading channel for spin relaxation
at fields below roughly $1$ Tesla. In addition to Ohmic fluctuations, the
quantum point contacts produce shot noise when driven out of
equilibrium, which further relaxes the spin~\cite{Bohrani05}.
Finally, $1/f$ background charge fluctuations, present in most
mesoscopic systems, also couple to the orbital motion of the
electrons and dephase the spin. In the following we shall develop a
formalism that allows us to treat different types of environments on
an equal footing, and also to take into account higher order virtual
processes, leading to a geometrical dephasing.

We consider a single electron confined to a lateral quantum dot by
the potential $V(\hat{\mathbf{r}})$, in the presence of a magnetic
field $\vec B$. To be specific, we assume the field to be oriented
parallel to the plane of the dot, but our procedure can be
generalized to arbitrary directions. The static part of the
Hamiltonian then reads
\begin{eqnarray}
H_{\rm 0} &=&\frac{\mathbf{p}^2}{2m^*}+V(\mathbf{r})-
\frac{g\mu_B}{2}\vec{B}\cdot\vec{\sigma}+H_{SO}\;, \\
H_{SO}&=&\alpha(p_y\sigma_x-p_x\sigma_y)+\beta
(p_y\sigma_y-p_x\sigma_x)\;.
\end{eqnarray}
The magnetic field couples to the electron only through a Zeeman
term with $g$-factor $g$ \cite{footnote0}. The last terms describe
the Dresselhaus ($\beta$) and Rashba couplings ($\alpha$) between
the spin $\vec \sigma$ of the electron and its momentum
\cite{Winkler03,footnote0}. For a dot of size $d$ the typical energy
of the spin-orbit coupling scales as $\sim \beta/d$, while the level
spacing scales as $\omega_0 \sim 1/(m^* d^2)$. Therefore, for dots
with small level spacing, $\omega_0 < 1 \;{\rm K}$, the spin-orbit
coupling cannot be treated perturbatively.

Next we account for time-dependent fluctuations of the
electromagnetic field, which add a term
\begin{equation}
\delta V (t) =  X^\mu(t) \; \hat{O}_\mu \label{V(t)}
\end{equation}
to the Hamiltonian. (A summation over repeated indexes, such as
$\mu$, is assumed throughout.) The  terms $\hat{O}_\mu$ denote
independent operators in the Hilbert space of the confined electron
(e.g., $x$, $y$, $x^2,\dots$), while the terms $X^\mu(t)$ denote the
corresponding fluctuating (in general quantum) fields (e.g., $\delta
E_x$, $\delta E_y$, $\nabla_x \delta E_x$, ...). They may be
generated by various environments, such as phonons, localized
defects, or electron-hole excitations. Information about their
specific properties is contained in the spectral functions, to be
specified later. Note that this formulation covers also quadrupolar
fluctuations.


\begin{figure}
\includegraphics[width=8.2 cm]{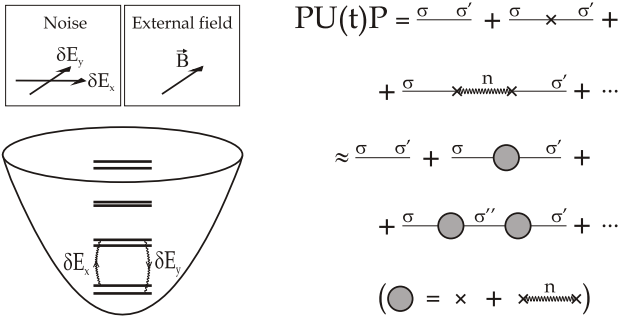}
\caption{ Left: Single electron lateral quantum dot in a magnetic
field, which lifts the ground state degeneracy.
Virtual transitions to excited states are induced by weak
fluctuations of the external fields $\delta E_x(t)$, $\delta
E_y(t)$. Right: Graphical representation of the evolution operator.
Virtual transitions to excited states $n$ (wavy lines) are
integrated out to yield an effective Hamiltonian within the doublet
subspace. }\label{fig:perturb}
\end{figure}

In case of time-reversal symmetry the ground state of the dot is
two-fold degenerate. This degeneracy is split in an external
magnetic field. If $g \mu_B B \ll \omega_0$, and as long as the
noise is {\it adiabatic} with respect to the orbital level
splitting, $T\ll \omega_0$, the dynamics of the spin remains
constrained to these two states. Under these conditions (following
the method described in Ref.~\onlinecite{Hutter06})
 we can derive an effective Hamiltonian for the
two lowest eigenstates $|\sigma= \pm\rangle$ by expanding the
evolution operator $U(t) = \mathrm{T\; exp}\{ -i\int_0^t dt'
\;{\delta V}_{\rm int}(t')\}$ and projecting to the subspace
$\{|\sigma\rangle\}$ as sketched in Fig.~\ref{fig:perturb}. This
yields
\begin{eqnarray}
&&PU(t)P  =   1-i\int_0^tdt_1P \delta V_{\rm int}(t_1) P\nn \\
&&-\int \int_{t_1> t_2}  P \delta V_{\rm int}(t_1) P \delta V_{\rm
int}(t_2)P
\label{eq:expansion} \\
&&-\int \int_{t_1> t_2}  P \delta V_{\rm int}(t_1) (1-P) \delta
V_{\rm int}(t_2)P +\ldots\; \nn .
\end{eqnarray}
Here $\delta V_{\rm int}(t)$ denotes the fluctuating part of the
Hamiltonian in the interaction representation and $P = \sum_\sigma
|\sigma\rangle\langle \sigma|$. We separated terms that involve
direct transitions between the two lowest states from transitions
via excited states. In the spirit of an adiabatic approximation,
these latter processes can be integrated out to yield an effective
Hamiltonian in the two-dimensional subspace. Technically, this is
performed by introducing  slow and fast variables, $t \equiv (t_1 +
t_2)/2$ and  $\tau \equiv t_1 - t_2$, in the last term of
Eq.~(\ref{eq:expansion}),
\begin{eqnarray}
\sim \;  e^{-i t(\epsilon_\sigma-\epsilon_{\sigma'}) -i\tau
\left[\frac{1}{2}(\epsilon_\sigma+\epsilon_{\sigma'})-\epsilon_n\right]}
\; \delta V_{\sigma n}(t_1) \delta V_{n\sigma'} (t_2)\;, \nonumber
\end{eqnarray}
expanding the interaction potential in $\tau $ as $ \delta
V(t_{1,2}) \approx \delta V(t) \pm
\frac{\tau}{2}\;\frac{d}{dt}{\delta  V}(t) +\dots$, and integrating
with respect to $\tau$. Here $\epsilon_\sigma$ and $\epsilon_n$
denote the eigenenergies of the lowest doublet and higher
eigenstates of $H_0$, respectively. In this way the last term in
Eq.~(\ref{eq:expansion}) becomes {\em local in time}. Retaining only
processes up to 2nd order, we find an effective Hamiltonian within
the lowest-energy two-dimensional subspace, characterized by the
'pseudospin' Pauli matrices  $\tau_{x,y,z}$,
\begin{eqnarray}
H_{\mathrm{eff}}& =& -\frac{1}{2}B_{\rm eff}\; {\tau_z} + X^\mu
\vec{C}^{(1)}_\mu\cdot\vec{\tau}  + X^\mu
X^\nu\vec{C}^{(2)}_{\mu\nu}\cdot\vec{\tau} \nn
\\
&+&\frac{1}{2}\left(\dot{X}^\mu
X^\nu-X^\mu\dot{X}^\nu\right)\vec{C}^{(3)}_{\mu\nu}\cdot\vec{\tau}
\; . \label{Heff}
\end{eqnarray}
Due to the spin-orbit coupling, which is not assumed to be weak,
eigenstates do not factorize into orbital and spin sectors (hence
the term 'pseudospin'). The static effective field, ${\vec B}_{\rm
eff} \equiv (\epsilon_{+} - \epsilon_{-})\hat{z}$, accounts for the
spin-orbit renormalization of the $g$-factor and defines the
$\hat{z}$ direction in the doublet space. The couplings $\vec
C^{(i)}$, determining the effective fluctuating magnetic fields felt
by the pseudospin, are given by
\begin{eqnarray}
&&\left[\vec{C}^{(1)}_\mu\cdot\vec{\tau}\right]_{\sigma,\sigma'}=\hat{O}^\mu_{\sigma
\sigma'}\ ,\\
&&\left[\vec{C}^{(2)}_{\mu\nu}\cdot\vec{\tau}\right]_{\sigma,\sigma'}=-
{\sum_n}^\prime \frac{\hat{O}^\mu_{\sigma n}\hat{O}^\nu_{n \sigma'}}
{\frac{\epsilon_\sigma+\epsilon_{\sigma'}}2-\epsilon_n}\ ,\\
&&\left[\vec{C}^{(3)}_{\mu\nu}\cdot\vec{\tau}\right]_{\sigma,\sigma'}=
-i\,{\sum_n}^\prime \frac{\hat{O}^\mu_{\sigma n}\hat{O}^\nu_{n
\sigma'} }
{\left(\frac{\epsilon_\sigma+\epsilon_{\sigma'}}{2}-\epsilon_n\right)^2}\
.
\end{eqnarray}
The summation is restricted to excited states of higher doublets
$n\ne \sigma,\sigma'$. We do not provide explicit expressions for
the eigenenergies $\epsilon_{\sigma}$, $\epsilon_{n}$, matrix
elements $\hat{O}^\mu_{\sigma \sigma'}$, $\hat{O}^\mu_{\sigma n}$,
or couplings $\vec C^{(i)}$, but below we will evaluate them
numerically and provide quantitative estimates for a generic model.
We further note that both $\vec{C}^{(1)}_\mu$ and
$\vec{C}^{(3)}_{\mu\nu}$ turn out to be transversal to ${\vec
B}_{\rm eff}$, therefore contributing only to relaxation, whereas
$\vec{C}^{(2)}_{\mu\nu}$ has in general also a parallel component
that leads to pure dephasing.

In time-reversal symmetric situation, (i.e. for $B=0$), the first
three terms of Eq.~(\ref{Heff}) vanish identically
\cite{Khaetskii01}. Only the last term survives, and leads to spin
dephasing. It has a {\em geometrical} origin. To demonstrate this,
let us assume that the fluctuating (adiabatic) fields $X_\mu$ are
classical. We introduce the instantaneous ground states of the
Hamiltonian, $|\Phi_n (t)\rangle \equiv |\Phi_n(X_\mu(t))\rangle$
defined through the equation \be[H_0+\delta V(X_\mu)] |\Phi_n(X_\mu)
\rangle = E_n(X_\mu) |\Phi_n(X_\mu) \rangle\;.\ee Noting that, to
lowest order perturbation theory, the two degenerate instantaneous
ground states are simply  given by $ |\Phi_\sigma(X_\mu) \rangle
\approx |\sigma\rangle + {\sum_n}^\prime |n\rangle {\langle n|\delta
V|\sigma\rangle} /({\epsilon_\sigma-\epsilon_n}) \;,$ we can rewrite
the last term in Eq.~(\ref{Heff}) in the  familiar form
\be H^{\mathrm{eff}}_{\sigma \sigma'}(B = 0) = -i \langle {\frac{d
\Phi_\sigma}{dt}  | \Phi_{\sigma'} }\rangle \;, \label{eq:berry}
\ee which shows clearly that the last term is due to a generalized
(possibly non-Abelian) Berry phase~\cite{Berry84,Wilczek84,Mead87}
acquired in a degenerate 2D subspace. In vanishing magnetic field,
Eq.~(\ref{eq:berry}) can be shown to hold to all orders of
perturbation theory within the adiabatic approximation. If at least
two linearly independent fluctuating fields couple to the dot, they
can produce a random Berry phase for the system and cause geometric
dephasing at $B=0$. When more noise components are present, the
Berry phase may become non-Abelian and all components of the spin
may decay.

So far, our treatment has been rather general, applicable for
arbitrary noise properties and dot geometries. In its full glory,
Eq.~(\ref{Heff}) describes the motion of the pseudo-spin coupled to
three fluctuating ``magnetic fields". In general, the dynamics
induced by these non-commuting fields is complicated. To obtain a
qualitative understanding of the dynamics we analyze the spin
relaxation and pure dephasing times~\cite{Bloch57}, $T_1$ and
$T_2^*$ (with $1/T_2 = 1/2T_1+1/T_2^*$). They are defined only for
sufficiently strong effective fields, $B_{\rm eff}\gg 1/T_1,
\;1/T^{*}_2$. In the limit $B=0$ we evaluate what we call the
geometrical dephasing time $T_{\rm geom}$. For the quantitative
estimate we consider a parabolic confining potential, $V({\bf r}) =
\frac{m^* \omega_0^2}{2} |{\bf r}|^2$ with level spacing $\omega_0$
and typical size $x_0=1/\sqrt{\omega_0 m^*}$.
Furthermore, we take  into account only dipolar fluctuations, $\hat
X \equiv e \delta E_x x_0$ and $\hat Y \equiv e \delta E_y x_0$
coupling to the operators ${\hat O}_X \equiv x/x_0$ and ${\hat O}_Y
\equiv y/x_0$, respectively. We assume the two components $\hat X $
and $\hat Y$ to be independent of each other, but with identical
noise spectra, $S_X(\omega) = S_Y(\omega) = S(\omega) = \pi
\varrho(\omega) {\rm coth}(\omega/2T)$, with  $\varrho(\omega)$
being the spectral function of the bosonic environment (phonons or
photons).

The spectral function $\varrho(\omega)$ for phonons can be estimated
along the lines of Ref.~\onlinecite{Khaetskii01}. For the parameters
specified in Ref.~\onlinecite{Golovach04} we find for piezoelectric
phonons in typical GaAs heterostructures at low frequencies,
$\varrho_{\rm ph} (\omega) = x_0^2 \; \lambda_{\rm ph} \; \omega^3$
with $\lambda_{\rm ph} \approx 4 \cdot 10^{-6} {\rm K}^{-2} {\rm nm}^{-2}$
\cite{footnote0}. With these parameters we obtain relaxation rates
generated by  the first term in Eq.~(\ref{Heff}) that coincide with
those of Ref.~\onlinecite{Golovach04} at not too high values of the
field \cite{footnote2}. Similar values are obtained for the
parameters of Ref.~\onlinecite{Stano05}. For Ohmic fluctuations the
spectral function is linear at low frequencies, $\varrho_\Omega
(\omega) = \lambda_\Omega \;\omega$ \cite{Leggett87,Weiss99}. The
prefactor $\lambda_\Omega$ depends on the
dimensionless impedance of the circuit,  $\lambda_\Omega\sim \frac { e^2} h \;{\rm Re }[Z]$. For typical values of the
sheet resistance of the 2-DEG ($10^2-10^3\Omega/\Box$) we estimate
it to be in the range $0.1 > \lambda_\Omega > 0.01$. For $1/f$ noise
the power spectrum is $S(\omega) = \lambda_{1/f} /|\omega|$. We will
further comment on its strength below.

We first estimate the contributions $T_1^{(i)}$ and $T_2^{*(i)}$,
derived from the three terms ($i=1,2,3$) in Eq.~(\ref{Heff}), for a
non-vanishing in-plane magnetic field, $B_{\rm eff}\gg 1/T_1,
\;1/T^{*}_2$. The coupling ${\vec C}^{(1)}$ turns out to be
 perpendicular to ${\vec B}_{\rm eff}$ \cite{Golovach04}, and
for low magnetic fields and weak spin-orbit coupling
 is proportional to $|{\vec C}^{(1)}|\sim
\frac{B}{x_0 \omega_0^2} \;\max\{{\alpha}, {\beta}\}$. This fluctuating
field therefore contributes to the $T_1$-relaxation only,
\be
\frac{1}{T_1^{(1)}} = 2 \left( |\vec{C}^{(1)}_{X}|^2 +
|\vec{C}^{(1)}_{Y}|^2\right) \;S_X(B_{\rm eff})\ .
\ee
It scales as $1/T_1^{(1)}\sim  B^2 \max \{B,T \}$ for Ohmic
dissipation and as $\sim  B^4 \max\{B,T\}$ for phonons. As a
consequence, for dots with level spacing in the range $\omega_0
\approx 1 \dots 10 \;{\rm K}$ Ohmic fluctuations dominate over
phonons for low fields with  $B < 1 \dots 3 \;{\rm T}$.

The second term, $|{\vec C}^{(2)}|\sim \frac{B}{x_0^2\omega_0^4}
\max\{\alpha ^2 , \beta ^2\}$ \footnote{Note that there is no
discrepancy with the scaling $\max\{\alpha, \beta\}$ in Ref. [5],
since they include multipolar spin-flipping phonon contributions
neglected here.} gives rise to both relaxation and dephasing. The
two rates
are
\bea
\frac{1}{T_1^{(2)}} =  4\left(|\vec{C}^{(2,\perp)}_{XX,s}|^2 +
|\vec{C}^{(2,\perp)}_{YY,s}|^2 +2
|\vec{C}^{(2,\perp)}_{XY,s}|^2\right) S_{XY}(B_{\rm eff})\;,
\nonumber
\\
\frac{1}{T_2^{*(2)}} =  4\left(|\vec{C}^{(2,\parallel)}_{XX,s}|^2 +
|\vec{C}^{(2,\parallel)}_{YY,s}|^2 +2
|\vec{C}^{(2,\parallel)}_{XY,s}|^2\right) S_{XY}(0)\;, \nonumber
\\
S_{XY}(\omega) = \frac{\pi}{2} \int d\tilde \omega\;
\frac{\varrho(\frac{ \omega + \tilde\omega}{2}) \varrho(\frac{\omega
- \tilde\omega}{2})} {1- \cosh(\tilde \omega/2T)/\cosh(\omega/2T) }\
, \nonumber
\eea
with $\vec{C}^{(2,\perp/\parallel)}_{\mu\nu,s}$ denoting the
symmetrized component of ${\vec C}^{(2)}_{\mu\nu}$
perpendicular/parallel to ${\vec B}_{\rm eff}$. Thus, for Ohmic
dissipation ${1}/{T_1^{(2)}}$ vanishes as $\sim  B^2 \max \{B^3,T^3
\}$, while for phonons it scales as $\sim  B^2 \max \{B^7,T^7 \}$.

$\vec{C}^{(3)}$ is also perpendicular to ${\vec B}_{\rm eff}$. Its
contribution to the relaxation is
\bea
\frac{1}{T_1^{(3)}} & = & 2 |\vec{C}^{(3)}_{XY,a}|^2
S_{\dot X Y - X\dot Y }(B_{\rm eff})\;,\\
\nonumber
 S_{\dot X Y - X\dot Y }(\omega) &=&\frac{\pi}{2} \int d\tilde \omega\;
\frac{\tilde\omega^2\varrho(\frac{ \omega + \tilde\omega}{2})
\varrho(\frac{\omega - \tilde\omega}{2})} {1- \cosh(\tilde
\omega/2T)/\cosh(\omega/2T) } \; ,
\eea
with $\vec{C}^{(3)}_{\mu\nu,a}$ being the anti-symmetrized component
of ${\vec C}^{(3)}_{\mu\nu}$. Most importantly, with $|{\vec
C}^{(3)}_{XY}|\sim \frac{1}{x_0^2\omega_0^4} \max\{\alpha ^2 ,\beta
^2\}$, the rate ${1}/{T_1^{(3)}}$  approaches a non-zero value at
low fields, $1/T_1,1/T_2^* \ll B\ll T$, and scales as $\sim \max
\{B^5,T^5 \}$ for Ohmic dissipation and $\sim \max \{B^9,T^9 \}$ for
phonons.

Finally,  at $B=0$ the geometric dephasing rate is given by
${1}/{T_1^{(3)}}$, extrapolated to zero field
\bea
\frac{1}{T_{\rm geom}} =  2 |\vec{C}^{(3)}_{XY,a}|^2 S_{\dot X Y -
X\dot Y }(\omega\sim 0)\ .
\eea
In our example with only two noise components this process dephases
only the components of the spin perpendicular to
$\vec{C}^{(3)}_{XY}$.

\begin{figure}
\includegraphics[width=8.6 cm]{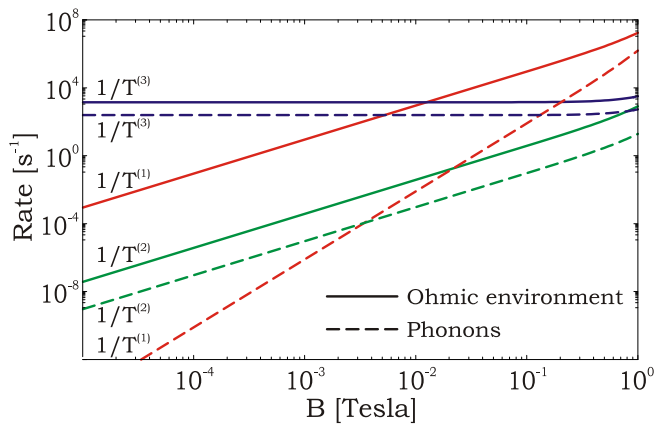}
\caption{\label{fig:rates} Spin relaxation rates  for a GaAs quantum
dot with level spacing $\omega_0 = 1{\rm K}$ as function of the
Zeeman field. We chose a temperature $T= 100 \;{\rm mK}$ and Ohmic
coupling $\lambda_\Omega = 0.05$. Below $B^*\approx 1\; {\rm T}$
spin relaxation is dominated by coupling to Ohmic fluctuations. For
$B < B^{**}\approx 15 \;{\rm mT}$ geometrical spin relaxation due to
coupling to Ohmic fluctuations dominates. For all  $B$ values
plotted, the Bloch-Redfield consistency requirement,
$B_{\mathrm{eff}}\gg 1/T^{(i)}$, is satisfied.}
\end{figure}

The relaxation rates corresponding to the different terms in
Eq.~(\ref{Heff}) and various noise sources  are shown in
Fig.~\ref{fig:rates}.
Clearly, for external fields $B< B^* \approx 1  \dots 3\;{\rm T}$
Ohmic fluctuations provide the leading relaxation mechanism. The
crossover field $B^*$ is not very sensitive to the specific value of
the spectral parameter $\lambda_\Omega$ and is independent of the
spin-orbit coupling. Below a second crossover field, $B^{**} \approx
15 {\rm mT}$, the geometric dephasing induced by Ohmic fluctuations
starts to dominate. This second crossover scale is very sensitive to
the spin-orbit coupling and temperature, scaling as $B^{**}\sim
\max\{\alpha,\beta\} (1/x_0) (T/\omega_0)^2$. E.g.\ for a level
spacing $\omega_0\sim 1{\rm K}$ and temperature $T=100 \;{\rm mK}$
($T=50 \;{\rm mK}$) the Berry phase mechanism gives a relaxation
time of the order of $700\;{\rm \mu s}$ ($20\;{\rm ms}$). For even
lower temperatures or  smaller dots with level spacing $\omega_0\sim
10 {\rm K}$ the $B\to 0$ relaxation time is quickly pushed up to the
range of  seconds.

Finally, we comment on the effect of $1/f$ noise. In most cases, the
non-symmetrized correlators for $1/f$ noise, needed to calculate
correlators as $S_{XY}$ or $S_{XX}$, are not known. Yet, for
$|\omega|\ll T$ we can provide an estimate $S_{XY}(\omega) \approx
\int\limits_{-T}^{T}\frac{d\tilde\omega}{2\pi} S_X(\omega -
\tilde\omega) S_Y(\tilde\omega)$, and similarly for
$S_{\dot{X}Y-X\dot{Y}}$. The $B=0$ geometrical dephasing rate due to
the $1/f$ noise can be estimated as $T_{\rm geom}^{-1}\approx|{\vec
C}^{(3)}_{XY}|^2 \lambda_{1/f}^2(T)\, \omega_c$, where $\omega_c$ is
the upper frequency cut-off for the $1/f$ noise~\cite{Ithier05}.
Accounting for the high-frequency (Ohmic) noise, sometimes observed
to be
 associated with the $1/f$
noise~\cite{Astafiev04,Shnirman05,Faoro05,Astafiev06}, the estimate
becomes $T_{\rm geom}^{-1}\approx|{\vec C}^{(3)}_{XY}|^2
\lambda_{1/f}^2(T)\, T$. While the $1/f$ noise of background charge
fluctuations is well studied in mesoscopic systems, the amplitude of
the $1/f$ noise of the electric field in quantum dot systems is yet
to be determined. If we assume that this noise is due to two-level
systems at the interfaces of the top gate electrodes, we conclude
that in the parameter range explored here, the effect of $1/f$ noise
is less important than that of Ohmic fluctuations. However, in
quantum dots with large level spacings in the low-temperature and
low-field regime, these fluctuations could dominate over the effect
of  Ohmic fluctuations and eventually determine the spin relaxation
time.

We thank J. Fabian for valuable discussions. This work has been
supported through a research network of the Landesstiftung
Baden-W\"urttemberg gGmbH, by Hungarian Grants OTKA Nos. NF061726,
T046267, and T046303, and the SQUBIT2 project IST-2001-39083.

\bibliographystyle{apsrev}
\bibliography{Berry_final}

\begin{thebibliography}{29}
\bibitem[{foo({\natexlab{a}})}]{footnote0}
\bibinfo{note}{For GaAs $m^*\approx 0.067\, m_e$, $g\approx -0.44$,
  longitudinal and transverse sound velocities $v_l= 4.73\cdot 10^3 m/s$,
  $v_t=3.35\cdot 10^3 m/s$, piezoelectric const.\ $h_{14}=1.4\cdot 10^9 eV/m$,
  density $\rho=5.3\cdot 10^3 Kg/m^3$,
  $\lambda_{ph}=(e^2h_{14}^2/105(2\pi)^2\rho (3/v_l^5+4/v_t^5)$. For the
  quantitative analysis we considered Dresselhaus coupling with
  $\lambda_{SO}=1/(m^*\beta)=1\mu m$.}
\bibitem[{foo({\natexlab{b}})}]{footnote2}
\bibinfo{note}{For larger frequencies the approximation $\varrho_{\rm ph}
  (\omega) = x_0^2 \; \lambda_{\Omega} \; \omega^3$ is not valid since the
  wavelength of relevant phonons becomes comparable to the size of the dot.}
\end{thebibliography}

\end{document}